\documentclass[conference, a4paper, 10pt]{IEEEtran}

\usepackage{graphicx,amssymb,amstext,amsmath,color}
\DeclareGraphicsExtensions{.eps,.pdf,.png,.jpg,.gif,.jpeg,.pstex}

\input{epsf.sty}

\usepackage{times, epsf}
\usepackage{amsmath,amssymb}
\usepackage{url}
\usepackage[font = footnotesize]{caption}
\usepackage{subfig}

\usepackage{float}
\usepackage{graphicx}
\usepackage{cite}
\usepackage{multirow,tabularx}
\usepackage[pdftex]{epsfig}
\input{epsf.sty}

\usepackage{stackrel}
\usepackage{amsfonts}
\input{epsf.sty}

\usepackage{times, epsf}
\usepackage{graphicx}
\usepackage{cite}
\usepackage{multirow,tabularx}
\usepackage{epstopdf}

\newcommand{\expect}[1]{{\mathbb{E}}\left[{#1}\right]}

\newcommand{\E}[1]{\expect{#1}}

\newcommand{\ba}{\begin{array}}
\newcommand{\ea}{\end{array}}

\def\PARstart#1#2{\begingroup\def\par{\endgraf\endgroup\lineskiplimit=0pt}
    \setbox2=\hbox{\uppercase{#2} }\newdimen\tmpht \tmpht \ht2
    \advance\tmpht by \baselineskip\font\hhuge=cmr10 at \tmpht
    \setbox1=\hbox{{\hhuge #1}}
    \count7=\tmpht \count8=\ht1\divide\count8 by 1000 \divide\count7 by\count8
    \tmpht=.001\tmpht\multiply\tmpht by \count7\font\hhuge=cmr10 at \tmpht
    \setbox1=\hbox{{\hhuge #1}} \noindent \hangindent1.05\wd1
    \hangafter=-2 {\hskip-\hangindent \lower1\ht1\hbox{\raise1.0\ht2\copy1}
    \kern-0\wd1}\copy2\lineskiplimit=-1000pt}

\newcommand{\bit}{\begin{itemize}}
\newcommand{\eit}{\end{itemize}}

\long\def\comment#1{}

\newfont{\bbb}{msbm10 scaled 700}

\newfont{\bb}{msbm10 scaled 1100}

\newcommand{\Dm}{{\bf D}}

\newcommand{\Gm}{{\bf G}}

\newcommand{\Qm}{{\bf Q}}
\newcommand{\Rm}{{\bf R}}

\newcommand{\Ec}{{\cal E}}

\newcommand{\Gc}{{\cal G}}
\newcommand{\Hc}{{\cal H}}

\newcommand{\Nc}{{\cal N}}

\newcommand{\Uc}{{\cal U}}

\newcommand{\gammav}{\hbox{\boldmath$\gamma$}}

\newcommand{\muv}{\hbox{\boldmath$\mu$}}

\newcommand{\Thetav}{\hbox{\boldmath$\Theta$}}

\renewcommand{\arg}{{\hbox{arg}}}

\newcommand{\transp}{{\sf T}}

\IEEEoverridecommandlockouts

\begin{document}
\title{Utility Optimal Scheduling and Admission Control for Adaptive Video Streaming in Small Cell  Networks}

\author{
\IEEEauthorblockN{D. Bethanabhotla, G. Caire, M. J. Neely}
\IEEEauthorblockA{Ming Hsieh Department of Electrical Engineering\\
University of Southern California\\
Email: { \{bethanab, caire, mjneely\}@usc.edu}
}
}
\maketitle
\begin{abstract}
We consider the jointly optimal design of a transmission scheduling and admission control policy 
for adaptive video streaming over small cell networks.
We formulate the problem as a dynamic network utility maximization 
and observe that it naturally decomposes into two subproblems: 
admission control and transmission scheduling. The resulting algorithms are simple and suitable for distributed implementation. 
The admission control decisions involve each user choosing the quality of the 
video chunk asked for download, based on the network congestion in its neighborhood. 
This form of admission control is compatible with the current video streaming technology  
based on the DASH protocol over TCP connections.  Through simulations, we evaluate the performance of the proposed algorithm under realistic assumptions for a small-cell network.
\end{abstract}

\section{Introduction}  \label{sec:into-dilip}

We consider the problem of joint transmission scheduling and congestion control for adaptive video streaming in a 
small cell network. 
We formulate a Network Utility Maximization (NUM) problem in the framework of 
Lyapunov Optimization, 
and derive algorithms for joint  transmission scheduling and congestion control inspired by the drift plus penalty 
approach~\cite{neely2010stochastic}.  

Excellent surveys on various problem formulations in the NUM framework can 
be found in~\cite{kelly2006mathematics,yi2008stochastic,chiang2007layering}.
Initial work in NUM focused on networks with static connectivity and time-invariant 
channels. One of the first applications of NUM was to show that Internet congestion control in TCP 
implicitly solves a NUM problem where the variant of TCP dictates the exact shape 
of the utility function \cite{kelly2006mathematics,chiang2007layering}. This framework has been extended to wireless ad-hoc networks with time 
varying channel conditions (see \cite{neely2005dynamic,yi2008stochastic}  and the references therein) and to peer-to-peer networks in \cite{neely2011utility}.
Recent work on adaptive video scheduling over wireless channels appears in \cite{bethanabhotla2012joint} by the authors, and in \cite{joseph2012jointly}. 

For the problem at hand, as a consequence of the NUM formulation, we obtain an elegant decomposition of the optimal solution into two separate subproblems, 
which interact through the queue lengths. 
The subproblems are solved by distributed dynamic policies 
requiring only local queue length information at each network node. 
The network is formed by {\em users}, which place video streaming requests and wish to download sequences of video chunks 
corresponding to the desired video files, and {\em helpers} (or femto base stations), which contain cached video files and serve the user requests through the 
wireless channel. The transmission scheduling decisions are carried out independently by the {\em helpers},
while the admission control decisions are carried out independently by the {\em users}. 
In  particular, we notice that the independent choices made by every user in deciding the quality of the video chunk 
that should be downloaded at any given time  is compatible with the current technology based on client-driven 
{\em Dynamic Adaptive Streaming over HTTP} (DASH) for video on demand (VoD) 
systems~\cite{sanchez2011idash,begen2011watching}. 

Motivated by realistic typical system parameters,  we assume that the time and frequency selective wireless channel fading coherence time $\times$ 
bandwidth product is  small with respect to the number of signal dimensions spanned by the transmission of a video chunk. 
This implies that the rate scheduling decisions in the transmission scheduling policy can make use of the 
``ergodic'' achievable rate region of the underlying physical layer. In contrast, other system parameters such as the distance dependent path loss
and the quality-rate tradeoff profile of the video file may evolve at the same time scale of the video chunks, yielding non-ergodic dynamics.
Also, we have to consider that a coded video file is formed by ``chunks'' (group of pictures) whose statistics may change with time. 
Correspondingly, variable bit rate (VBR) video coding \cite{ortega2000variable} yields a quality-rate tradeoff profile that may 
change with time (i.e., with the chunk index). We address these non-stationary and non-ergodic 
dynamics of the system parameters by providing performance guarantees for {\em arbitrary sample paths}, 
using the approach developed in~\cite{neely2010stochastic,neely2010universal}.

\section{System Model} \label{sec:sysmodel-dilip}

We consider a discrete, time-slotted wireless network 
with multiple user stations and multiple helper stations. The network is defined by a bipartite graph
$\Gc = (\Uc, \Hc, \Ec)$, where $\Uc$ denotes the set of users, $\Hc$ denotes the set of helpers, and 
$\Ec$ contains edges for all pairs $(h,u)$ such that there exists a potential transmission link between $h \in \Hc$ and $u \in \Uc$. 
We denote by $\Nc(u) \subseteq \Hc$ the neighborhood of user $u$, i.e., 
$\Nc(u) = \{ h \in \Hc : (h,u) \in \Ec\}$. Similarly,  $\Nc(h) = \{u \in \Uc : (h,u) \in \Ec\}$. Each user $u \in \Uc$ requests a video file $f_u$ from a library of possible files. A video file is formed by a sequence of 
``chunks'', i.e., group of pictures (GOPs), that are encoded and decoded as stand-alone units.  
Chunks have the same duration in time, given by $T_{\rm gop} = \mbox{(\# frames per GOP)}/\eta$, where $\eta$ is the frame rate (frames per second). 
Chunks must be reproduced in sequence at the user end. The streaming process consists of transferring chunks from the helpers to the requesting users
such that the playback buffer at each user contains the required chunks at the beginning of each chunk 
playback time.  Streaming is different from downloading because the playback starts while the whole file has not been entirely transferred.
In fact, the playback starts after a short pre-fetching time, where the playback buffer is filled in by a determined amount of chunks.
We assume that the scheduler time-scale coincides with the chunk interval, i.e., at each chunk interval a scheduling decision is made.
Conventionally, we assume a slotted time axis $t = 0,1,2,3 \ldots, $ corresponding to epochs $t \times T_{\rm gop}$. 
Let $T_u$ denote the pre-fetching delay of user $u$. Then, chunks are downloaded starting at time $t = 0$ and playback starts at time
$T_u$. A stall event for user $u$ at time $t \geq T_u$ is defined as the event that the playback buffer does not contain 
chunk number $t - T_u$ at slot time $t$.

Helpers have caches that contain subsets of the video files in the library. 
We denote by $\Hc(f)$ the set of helpers that contain file $f$. Hence, the request of user $u$ for a chunk at a particular slot $t$ 
can be assigned to any one of the helpers in  the set $\Nc(u) \cap \Hc(f_u)$. 
Letting  $N_{\mathrm{pix}}$ denote the number of pixels per frame, a chunk contains $k = \eta T_{\rm gop} N_{\mathrm{pix}}$ 
pixels (source symbols). We assume that each chunk of each file $f$ is encoded at a finite number of different quality modes
$m \in \{1, \ldots, N_f\}$ which is similar to what is typically done in several recent video streaming technologies like Microsoft Smooth Streaming 
and Apple HTTP Live Streaming~\cite{begen2011watching}. Due to the variable bit-rate nature of video coding, the quality-rate profile 
may vary from chunk to chunk. We let $D_f(m,t)$ and $kB_f(m,t)$ denote the video quality measure and the number of bits for 
file $f$ at chunk time $t$ and quality mode 
$m$ respectively. A fundamental function of the network controller at every slot time $t$ consists of choosing the quality
mode $m_u(t)$ of the chunks requested at $t$ by each user $u$. The choice $m_u(t)$ renders the choice of the point $(D_{f_u}(m_u(t),t), 
kB_{f_u}(m_u(t),t))$ from the finite set of quality-rate tradeoff points $\{(D_{f_u}(m,t), 
kB_{f_u}(m,t))\}_{m=1}^{N_f}$. We let $R_{hu}(t)$ denote the source coding rate (bit per pixel) of chunk $t$ received 
by user $u$ from helper $h$. In addition to choosing the quality mode $m_u(t)$ for chunk time $t$ for all requesting users 
$u$, the network controller also allocates the source coding rates $R_{hu}(t)$ satisfying
\begin{align}
\label{downloadconst}
\sum_{h \in \Nc(u) \cap \Hc(f_u)}R_{hu}(t) = B_{f_u}(m_u(t),t)~\forall~(h, u) \in \Ec.
\end{align}
When $R_{hu}(t)$ is determined, helper $h$ places the corresponding $k R_{hu}(t)$ bits 
in its transmission queue $Q_{hu}$, to be sent to user $u$ within the queuing and transmission delays. Notice that in order to be able to download different parts of the same chunk from different helpers,
the network controller needs to ensure that all received
bits from the serving helpers $\Nc(u) \cap \Hc(f_u)$ are useful, i.e., the union of all received bits yields
the entire chunk, without overlaps and without gaps. However, in Section~\ref{sec:numiidstate-dilip}, we
 will see that even if we allow the possibility of downloading different parts of the same chunk from different helpers, the
 optimal scheduling policy is such that users download an entire chunk from a single helper, rather than obtaining different parts from different helpers. 
Thus, it turns out that the assumption of protocol coordination to prevent overlap or gaps of the 
downloaded bits from different helpers is not needed. 
The dynamics of the transmission queues at the helpers is given by:
\begin{align}
  Q_{hu}(t+1)=\max\{Q_{hu}(t)-n\mu_{hu}(t),0\}&+kR_{hu}(t) \notag \\
& \forall~ (h,u) \in \Ec
\end{align}
where $n$ denotes the number of physical layer channel symbols corresponding to the duration $T_{\rm gop}$,  
and $\mu_{hu}(t)$ is the channel coding rate (bits/channel symbol) of the 
transmission from helper $h$ to user $u$. 
We model the point-to-point wireless channel for each $(h,u) \in \Ec$ as a frequency and time selective 
underspread~\cite{tse-viswanath} fading channel.
Using OFDM, the channel can be converted in a set of $N_c$ 
parallel narrowband sub-channels in the frequency domain (subcarriers), each of which is time-selective 
with a certain fading channel coherence time. We assume the widely adopted 
block fading model, where the small scale Rayleigh fading coefficient is constant over 
time-frequency ``tiles'' (resource blocks) spanning blocks of adjacent subcarriers in the frequency domain and blocks of OFDM symbols in the time domain. 
For example, in the LTE 4G standard, for an available system bandwidth of $18\mathrm{MHz}$
(after excluding the guard bands) and a scheduling slot of duration $T_{\rm gop} = 0.5$s (typical GOP duration), 
we have that a scheduling slots spans $100\times 1000$ such resource blocks, each of which is affected by its own fading coefficient. Thus, it is safe to assume that channel coding over such a large
number of resource blocks can achieve the {\em ergodic capacity} of the underlying fading channel.~\footnote{This is the 
capacity averaged with respect to the first-order fading distribution, which has the operational meaning of an achievable rate only if coding across an arbitrarily 
large number of fading states is possible. In contrast, the non-ergodic ``outage capacity'' of the fading channel is relevant when a channel codeword spans a limited number of fading states, 
that does not increase with the channel coding block length.}
For simplicity, in this paper we consider constant power transmission, i.e., the serving helpers transmit with constant and flat power spectral density over the whole
system bandwidth, irrespective of the scheduling decisions and of the instantaneous fading channel state.
We further assume that every user $u$ when decoding a  transmission from a particular helper $h \in \Nc(u)$ treats the interference from 
other helpers as noise.  Under these system assumptions, the maximum achievable rate 
for link $(h,u) \in \Ec$ is given by 
\begin{equation}  \label{rateconst2}
 C_{hu}(t) = \E{ \log\left(1+\frac{P_hg_{hu}(t)|a_{hu}|^2}{1+\sum_{h^{'}\neq h}
  P_{h^{'}}g_{h^{'}u}(t)|a_{h^{'}u}|^2}\right)},
\end{equation} 
where $P_h$ is the transmit power of helper $h$, $a_{hu}$ is the small-scale fading gain from 
helper $h$ to user $u$ and $g_{hu}$ is the slow fading 
gain (pathloss) from helper $h$ to user $u$. 
We assume that each helper $h$ serves its neighboring users $u \in \Nc(h)$ using orthogonal FDMA/TDMA. 
Therefore, the set of rates $\{\mu_{hu}(t) : u \in \Nc(h)\}$ is constrained to be in the ``time-sharing region'' of  the broadcast channel formed
by helper $h$ and its neighbors $\Nc(h)$. This yields the transmission rate constraint
\begin{equation}  \label{rateconst1}
\sum_{u \in \Nc(h)} \frac{\mu_{hu}(t)}{C_{hu}(t)} \leq 1~ \forall~h\in \Hc.
\end{equation}
The slow fading gain $g_{hu}(t)$ models  path loss and shadowing between helper $h$ and user $u$,
and it is assumed to change very slowly with time. We let $\omega(t)$ denote the network state at time $t$, i.e., 
\[ \omega(t) = \left \{ g_{hu}(t), \left(D_{f_u}(\cdot, t), B_{f_u}(\cdot, t)\right) : \forall \; (h,u) \in \Ec,~u \in \Uc \right \}. \]

Let $A_{\omega(t)}$ be the set of feasible control actions, dependent on the current network state 
$\omega(t)$, and let $\alpha(t) \in A_{\omega(t)}$ be a control action, comprising the vectors 
$\Rm(t)$ with elements $k R_{hu}(t)$ of video coded bits, 
$\muv(t)$ with elements $n\mu_{hu}(t)$ of channel coded bits and the quality modes $m_u(t)~\forall~u \in \Uc$. A control policy is a sequence of control actions 
$\{\alpha(t)\}_{t=0}^{\infty}$ where at each time $t$, $\alpha(t) \in A_{\omega(t)}$.

\section{Problem Formulation and Control Policy Design} \label{sec:numiidstate-dilip}

We now formulate the Network Utility Maximization problem. The goal is to design a control policy which maximizes a concave 
utility function of the time averaged qualities of all users subject to keeping the queues at every helper stable. Define
$\overline{D}_u:=\lim_{t\rightarrow \infty}\frac{1}{t}\sum_{\tau=0}^{t-1}\E{ D_{f_u}\left(m_u(\tau),\tau\right)}$ as the time averaged 
quality of user $u$ and let $\phi_u(\cdot)$ be the concave, continuous, non-negative and non-decreasing utility function for
each user $u$. The goal is to solve:
\begin{align}
\textrm{max } &\sum_{u \in \Uc}\phi_u(\overline{D}_u)\label{maxutil}\\
 \textrm{subject to } &\lim_{t\rightarrow \infty}\frac{1}{t}\sum_{\tau=0}^{t-1}
\E{Q_{hu}\left(\tau\right)} < \infty~\forall~ (h,u) \in \Ec \label{qstableconst}\\
& ~\alpha(t) \in A_{\omega(t)}~\forall~t \label{feasibleoptions}
\end{align}
where constraint (\ref{qstableconst}) corresponds to the {\it strong stability} condition for all the queues $Q_{hu}$.
The above problem is solved using the stochastic optimization theory of~\cite{neely2010stochastic}.  Since it involves maximizing a {\it function} of time averages, it is first transformed, using auxiliary variables $\gamma_u(t)$, to the following  problem that involves maximizing a single time average instead of a function of time averages so that the {\it drift plus penalty} framework of~\cite{neely2010stochastic} can be applied :
\begin{align}
 \textrm{max } &\sum_{u \in \Uc}\overline{\phi_u({\gamma}_u)}\label{maxutiltrans}\\
 \textrm{subject to } &\lim_{t\rightarrow \infty}\frac{1}{t}\sum_{\tau=0}^{t-1}
\E{Q_{hu}\left(\tau\right)} < \infty~\forall~ (h,u) \in \Ec \label{qstableconsttrans}\\
&~\overline{\gamma}_u \leq \overline{D}_u~\forall~u~\in~\Uc \label{gammaconst}\\
&~D_u^{\min} \leq \gamma_u(t) \leq D_u^{\max}~\forall~u~\in~\Uc \label{rectconst}\\
& ~\alpha(t) \in A_{\omega(t)}~\forall~t \label{feasibleoptionstrans}
\end{align}
where $\overline{\gamma}_u$ is the time averaged expectation of $\gamma_u(t)$, $D_u^{\max}$ is a uniform upper bound on the maximum quality $D_{f_u}(N_{f_u}, t)$  and $D_u^{\min}$ is a uniform lower bound on the minimum quality $D_{f_u}(1,t)$ for all chunk times $t$. 
To satisfy constraints~(\ref{gammaconst}), for each $u \in \Uc$, we define the virtual queue:
\begin{align}
 \Theta_u(t+1) = \max{\{\Theta_u(t)+\gamma_u(t)-D_{f_u}(m_u(t),t),0 \}} \label{concvirt}
\end{align}
Notice that constraints~(\ref{gammaconst}) correspond to stability of the virtual queues $\Theta_u$, since
$\overline{\gamma}_u$ and $\overline{D}_u$ are the time-averaged arrival rate and the time-averaged 
service rate for the virtual queue given in (\ref{concvirt}). It is easily shown in~\cite{bethanabhotla2013joint} that the optimal utility value $\phi_{\mathrm{opt}}$ is the same for both problems (\ref{maxutil})-(\ref{feasibleoptions}) and (\ref{maxutiltrans})-(\ref{feasibleoptionstrans}).

Let ${\bf Q}(t)$,  ${\Thetav}(t)$ denote the column vectors of the queues $Q_{hu}(t)~\forall~(h,u)\in\Ec$,   virtual queues $\Theta_u(t)~\forall~u \in\Uc$ respectively. Also let $\gammav(t)$, $\Dm(t)$
denote the vectors with elements $\gamma_u(t) ~\forall~u \in\Uc$, $D_{f_u}(m_u(t),t)~\forall~u\in\Uc$ respectively. 
Let ${\bf G}(t)=\left[ {\bf Q}^\transp(t), \Thetav^\transp(t)\right]^\transp$
and define the quadratic Lyapunov function $L({\bf G}(t)) := \frac{1}{2} \Gm^\transp (t) \Gm(t)$. 
Defining $\Delta(t)=\E{L(t+1)|\Qm(t)}-L(t)$ as the drift
at slot $t$, the {\em drift plus penalty} (DPP) policy~\cite{neely2010stochastic} is designed to solve 
(\ref{maxutiltrans})-(\ref{feasibleoptionstrans}) by observing only the current queue lengths ${\bf Q}(t)$ and 
the current network state $\omega(t)$ on each slot $t$ and then choosing $\alpha(t) \in A_{\omega(t)}$ to minimize a bound 
on
\begin{equation*}
 \Delta(t) - VD(t). 
\end{equation*}
 Here, $V>0$ is a control parameter of the DPP
policy which affects a utility-backlog tradeoff.
It is shown in~\cite{bethanabhotla2013joint} that the above minimization reduces to:  minimize
\begin{align}
\underbrace{{\bf R}^\transp (t) {\bf Q}(t) - \Dm^\transp(t)\Thetav(t)}_{\mbox{\footnotesize admission control}} \;&-\;
\underbrace{{\boldsymbol \mu}^\transp (t){\bf Q}(t)}_{\begin{array}{c} \mbox{\footnotesize transmission} \\ [0.05 mm]\mbox{\footnotesize scheduling}\end{array}} \notag \\
& - \underbrace{\left [  V\sum_{u \in \Uc}\phi_u(\gamma_u(t)) \gammav^\transp(t)\Thetav(t)  \right ]}_{\mbox{\footnotesize obj. maximization}}
 \label{DDP}
\end{align}
for every slot $t$ using only the knowledge of ${\bf Q}(t)$ and $\omega(t)$. The choice of ${\bf R}(t)$ and $m_u(t)~\forall~u \in U$ affects 
only the term ${\bf R}^\transp (t) {\bf Q}(t) -\Dm^\transp(t)\Thetav(t)$, while the choice of
${\boldsymbol \mu}(t)$ affects only the term $ - {\boldsymbol \mu}^\transp (t) {\bf Q}(t)$ and the choice of $\gammav(t)$ 
affects only the term $\gammav^\transp(t)\Thetav(t)-V\sum_{u \in \Uc}\phi_u(\gamma_u(t))$.
Thus, the overall minimization decomposes into three separate minimizations.

\subsection{Admission Control}\label{subsec:no drop-adm control}
The admission control sub-problem involves minimizing the objective function ${\bf R}^\transp (t) {\bf Q}(t) - \Dm^\transp(t)\Thetav(t)$ (see (\ref{DDP})).
The minimization of this quantity decomposes into separate minimizations for each user, namely, for each $u \in \Uc$, choose $m_u(t)$ and $R_{hu}(t)~\forall~h \in \Nc(u) \cap \Hc(f_u)$ to minimize
\begin{align}
\sum_{h \in \Nc(u) \cap \Hc(f_u)} k Q_{hu}(t)R_{hu}(t) - \Theta_u(t) D_{f_u}\left(m_u(t),t\right)
\end{align}
with $\{R_{hu}(t)\}_{h \in \Nc(u)\cap \Hc(f_u)}$ satisfying~(\ref{downloadconst}).
It is immediate to see that the above problem is solved by choosing the helper $h^*_u \in \Nc(u)\cap \Hc(f_u)$ with the smallest queue backlog $Q_{hu}(t)$, 
and assigning the entire requested chunk to $h^*_u$. Notice that in this way the streaming of the video file $f_u$ may be handled by {\em different} helpers across
the streaming session, but each individual chunk is downloaded from a single helper.  Further, the quality mode 
$m_u(t)$ is chosen as
\begin{align} \label{source-coding-rate-decision}
\arg \min_{m \in \{1, \ldots, N_{f_u}\}}\{k Q_{h^*_u u}(t) & B_{f_u}(m,t) -\Theta_u(t)D_{f_u}(m,t)\}.
\end{align}
In order to implement this policy, it is sufficient that each user knows only its {\em local information} of the queue backlogs of its neighboring helpers.
This policy is reminiscent of the current  adaptive streaming technology for video on demand systems, 
referred to as DASH~\cite{sanchez2011idash}, 
where the client (user) progressively fetches a video file by downloading successive chunks, 
and makes adaptive decisions on the source encoding quality based on its current knowledge of the congestion of the 
underlying server-client connection. 
\subsection{Transmission Scheduling}
\label{subsec:no drop-scheduling}
Transmission scheduling involves maximizing the weighted sum rate 
$\sum_{h \in \Hc} \sum_{u \in \Nc(h)} Q_{hu}(t) \mu_{hu}(t)$ where the weights are the queue backlogs (see (\ref{DDP})). Under our system assumptions,  this problem decouples into separate maximizations for each helper. Thus, for each $h \in \Hc$, the transmission scheduling problem can be written as the {\em Linear Program} (LP):
\begin{align}
 \textrm{maximize} & \;\;\; \sum_{u \in \Nc(h)} Q_{hu}(t) \mu_{hu}(t)\\
 \textrm{subject to} & \;\;\; \sum_{u \in \Nc(h)}\frac{\mu_{hu}(t)}{C_{hu}(t)} \leq 1. 
\end{align} 
The feasible region of the above LP is the $|\Nc(h)|$-simplex polytope and it is immediate to see 
that the solution consists of scheduling the user $u^*_h \in \Nc(h)$ with the largest product $Q_{hu}(t) C_{hu}(t)$, and serve this user
at rate $\mu_{h u^*_h}(t) = C_{h u^*_h}(t)$, while all other queues of helper $h$ are not served in slot $t$. 
\subsection{Greedy maximization of the network utility function}
Each user $u \in \Uc$ keeps track of $\Theta_u(t)$ and chooses its virtual queue arrival $\gamma_u(t)$ in order to solve:
\begin{align}
\textrm{maximize}& \;\;\; V\phi_u(\gamma_u(t))-\Theta_u(t)\gamma_u(t) \\
\textrm{subject to }& \;\;\; D_u^{\min} \leq \gamma_u(t) \leq D_u^{\max}.
\end{align}
These decisions push the system to approach the maximum of the network utility function. 

\section{Algorithm Performance}
It is shown in~\cite{bethanabhotla2013joint}  that the time average utility achieved by the DPP policy comes within $O(\frac{1}{V})$ of the utility of a 
genie-aided $T$-slot look ahead policy for any arbitrary sample path $\omega(t)$ with a $O(V)$ tradeoff in time averaged backlog. The details are omitted due to space restrictions and can be found in \cite{bethanabhotla2013joint}.




\section{Numerical Experiment} \label{sec:simul-dilip}

\begin{figure}
\centering
\includegraphics[width = 50mm, height = 32mm]{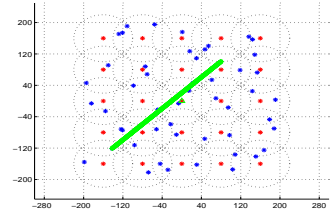}
\caption{Toplogy (the green line indicates the trajectory of a mobile user).}
\label{topology}
\end{figure}

We consider a $400$m $\times 400$m square area divided into $5\times5$ small square cells of side length $80$m as shown in Figure~\ref{topology}. 
A helper is located at the center of each small square cell. 
Each helper serves only those users within a radius of $60$ m.  
As described in Section \ref{sec:into-dilip}, the helpers could be connected to some video content delivery network through 
a wired backbone or they could be dedicated nodes with local caching capacity. In these simulations we assume that 
each helper has available the whole video library. Therefore, for any request $f_u$ we have
$\Nc(u) \cap \Hc(f_u) = \Nc(u)$. We further assume that there are $2$ users uniformly and independently distributed 
in each small cell. We use the utility function $\phi_u(x)=\log(x)$ (corresponding to proportional fairness) for all $u \in \Uc$.
We assume a physical layer inspired by LTE specifications~\cite{lte}. 
%
 
Between any two points $a$ and $b$ in the square area, the path loss is given by
$g(a,b) = \frac{1}{1+\left(\frac{d(a,b)}{\delta}\right)^\alpha}$
where $\delta = 40$ m and $\alpha=3.5$. 
Each helper transmits at a power level such that the SNR per symbol (without interference) 
at the center (i.e., at distance $d(a,b) = 0$ from the transmitter) is $20$ dB. 


\begin{figure}
\subfloat[bitrate profile]{\includegraphics[width = 43 mm, height = 35 mm]{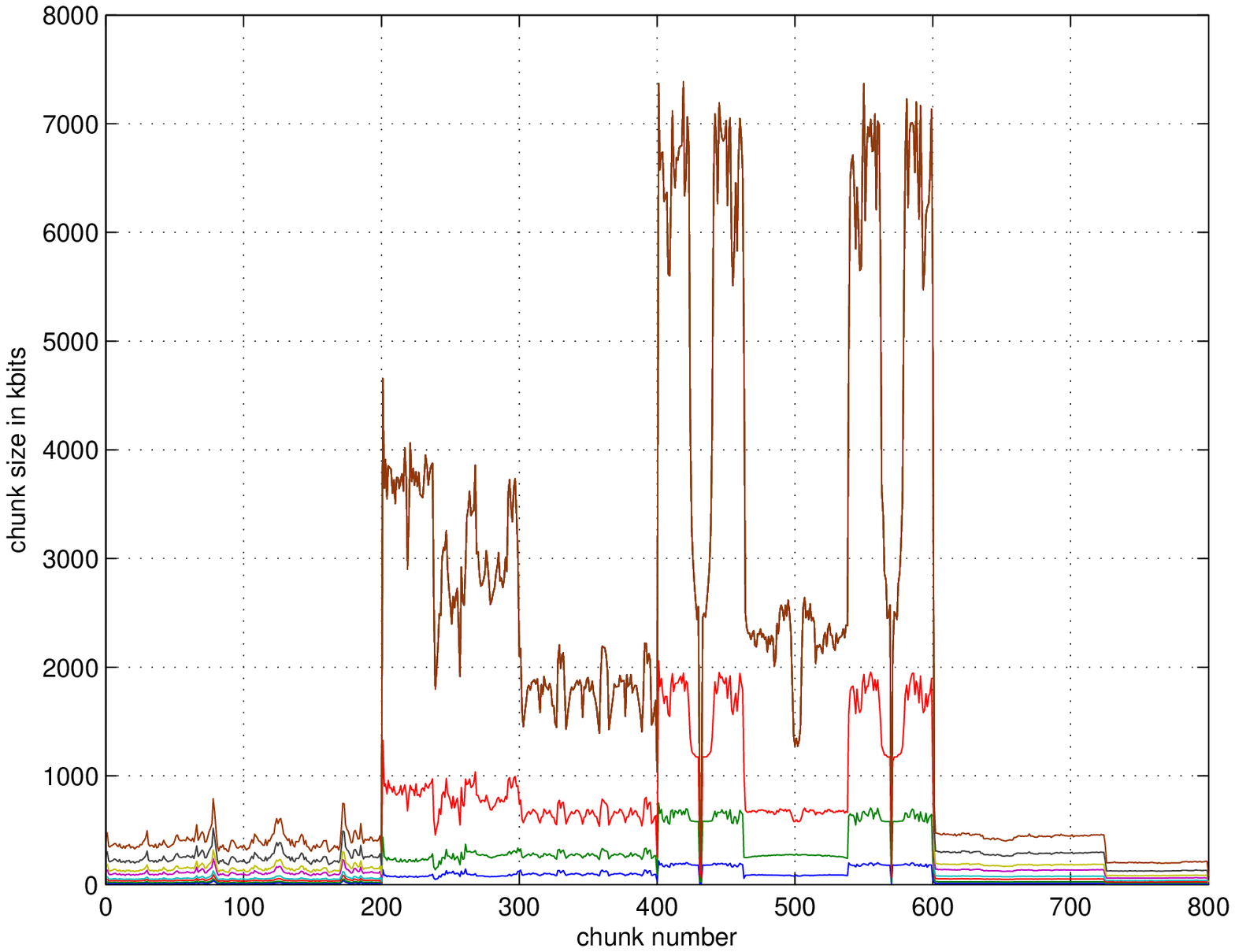}
\label{bitperchunk}} 
\subfloat[Quality profile]{
\includegraphics[width = 43 mm, height = 35 mm]{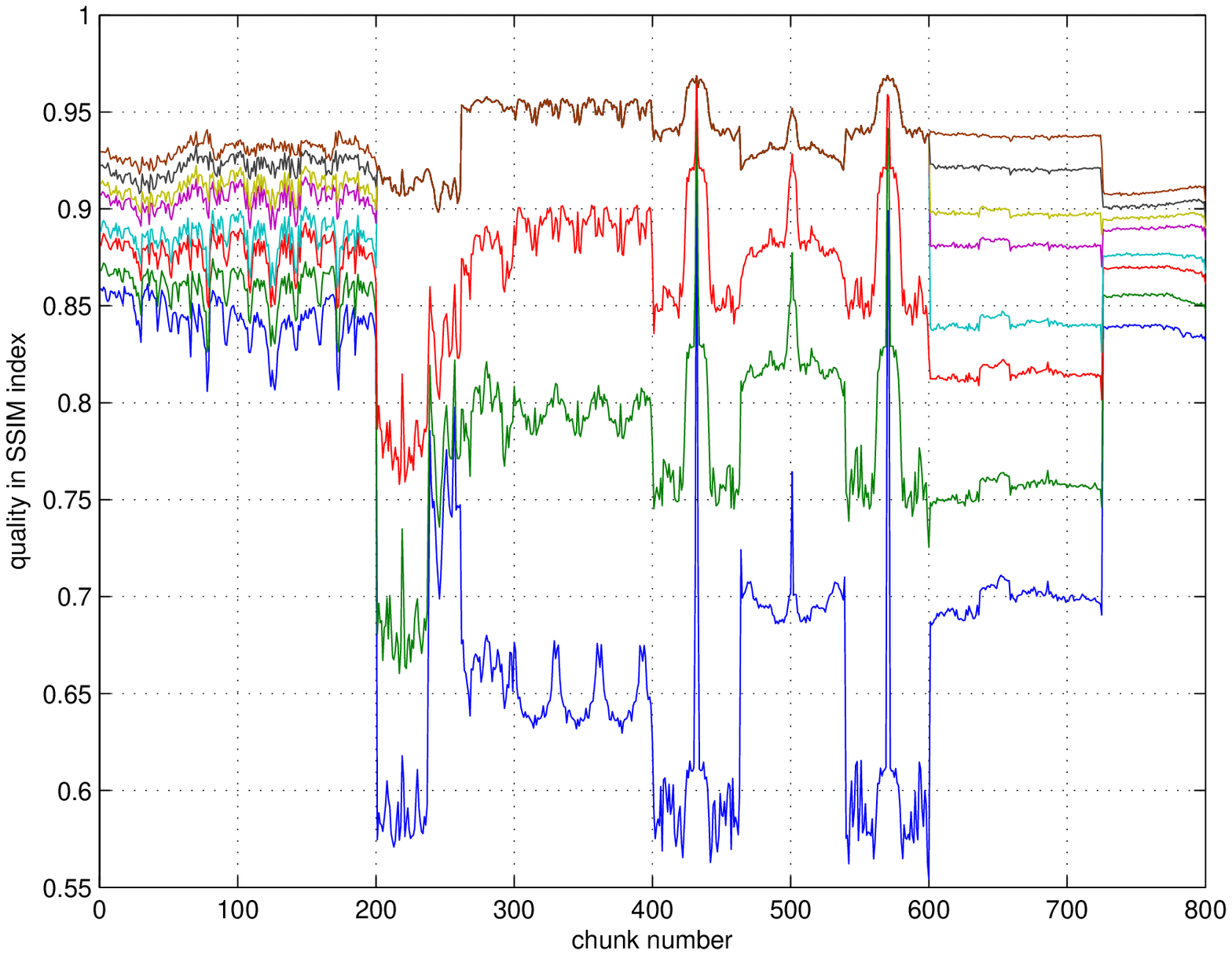}\label{ssimperchunk} 
}
\caption{Rate-quality profile of video sequence.}
\end{figure}

\begin{figure}
\centering
\includegraphics[width = 50mm, height = 35mm]{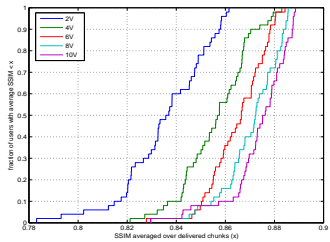}
\caption{CDF of quality over user population}
\label{ssim-cdf}
\end{figure}

\begin{figure}
\subfloat[Seamless downloading of chunks.]{
\includegraphics[width = 42 mm, height = 35mm]{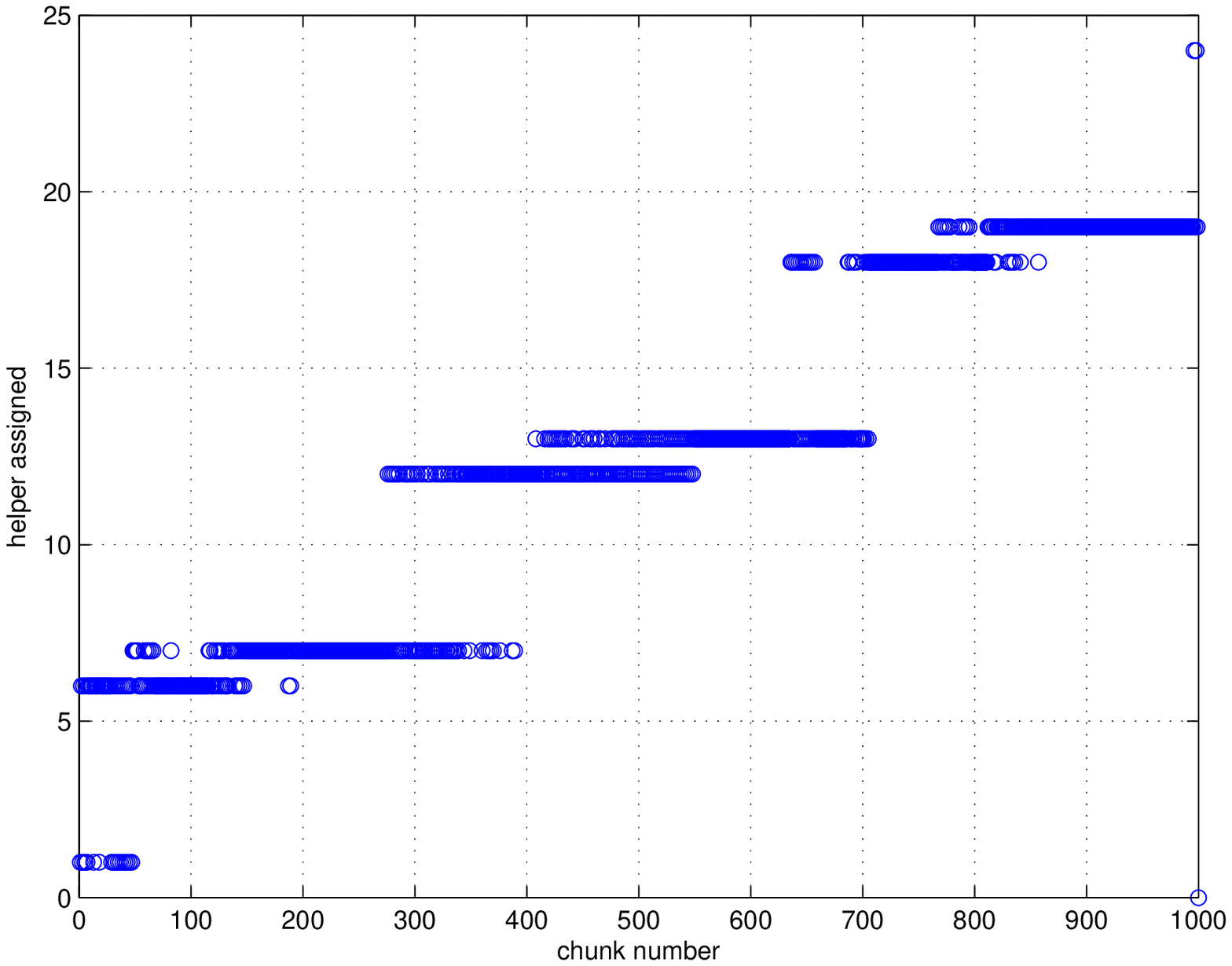}
\label{seamless}
}
\subfloat[Playback buffer dynamics.]{
\includegraphics[width = 42 mm, height = 35mm]{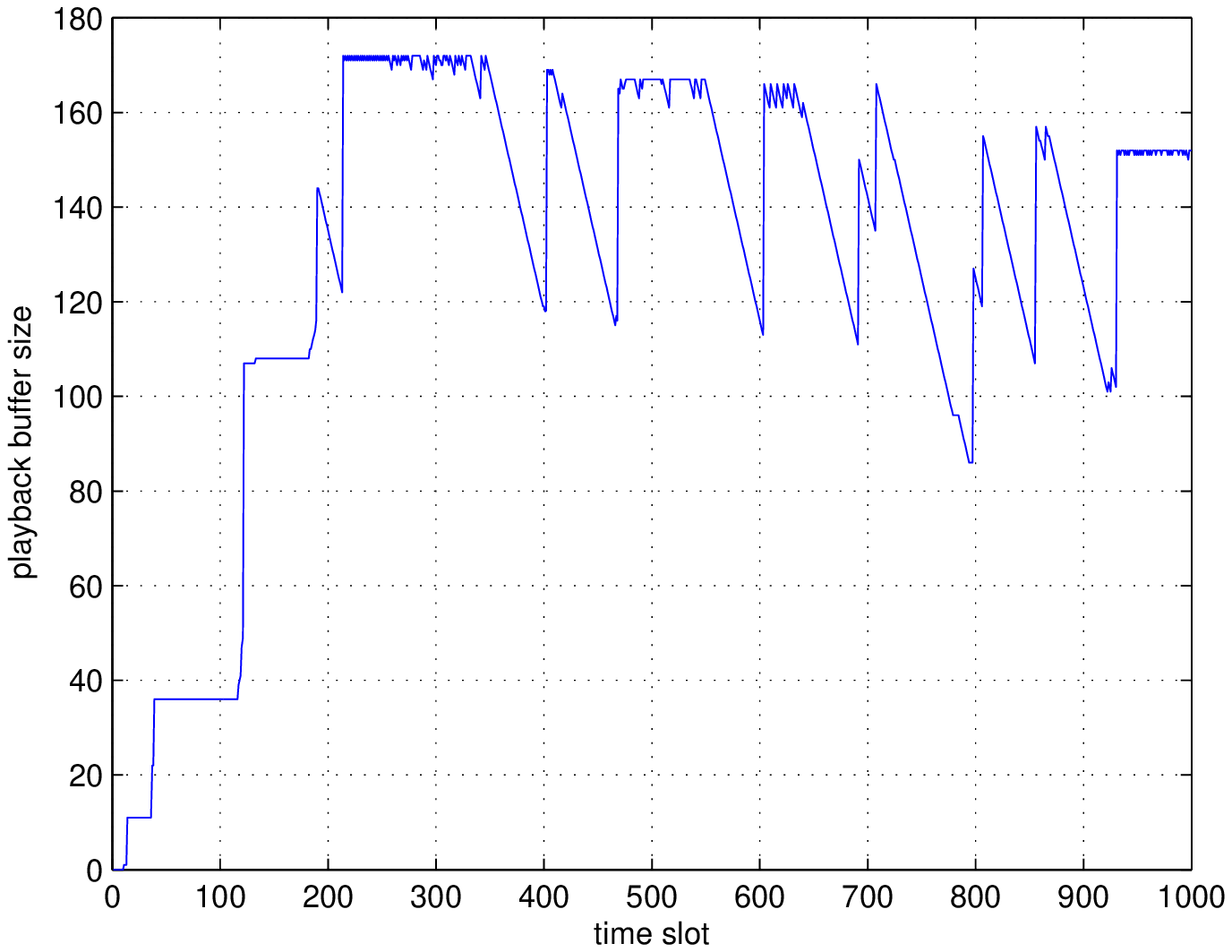}
\label{mobile-streaming}
}
\caption{Streaming performance of mobile user}
\end{figure}

We assume that all the users request chunks successively from VBR-encoded video sequences.  
Each video file is a long sequence of chunks, each of duration $0.5$ seconds 
and with a frame rate of $30$ frames per second. 
We consider a specific video sequence formed by $800$ chunks, constructed using $4$ video clips 
from the database in~\cite{video-samples}, each of length $200$ chunks. 
The chunks  are encoded into different quality modes. Here, the quality index is measured using the {\em Structural SIMilarity} (SSIM) index defined in \cite{ssim}.
Figures~\ref{bitperchunk} and \ref{ssimperchunk} show the size in kbits and the SSIM values as a function of the chunk index, respectively, 
for the different quality modes. The chunks from $1$ to $200$ and $601$ to $800$ are encoded into $8$ quality modes, 
while the chunks numbered from $201$ to $600$ are encoded in $4$ quality modes. 
In our experiment, each user starts its streaming session of $1000$ chunks from some arbitrary position in this reference video sequence and successively requests $1000$ chunks by cycling through the sequence. In addition, each user implements a policy to locally estimate the delay with which the video chunks are delivered, such that it can decide its pre-buffering time at the beginning of a streaming session or re-buffering time in the case of a ÒstallÓ event (empty playback buffer) during a streaming session. In addition, it may happen that chunks which go through different queues in the network are affected by different delays. This may give rise to a situation where already received chunks with higher order number cannot be used for playback until the missing chunks with lower order number are also received. In such a case, the policy also provides each user the flexibility to skip a chunk if by doing so, it can provide a large jump in its playback buffer. The complete details of this adaptive playback buffer policy are given in \cite{bethanabhotla2013joint}. Figure \ref{ssim-cdf} shows the cumulative distribution function (CDF) of quality (averaged over delivered chunks)
over the user population for the values $2V, 4V, 6V, 8V$ and $10V$ with $V = 10^{12}$. We can notice the fairness in service as the policy achieves a value close to the optimum for large $V$. 
We repeat the experiment with the same setup, but now we consider a specific user, indicated by $u_1$, moving slowly across the square grid along the green path indicated in Figure~\ref{topology}, during the $1000$ slots of simulation. 
We fix the parameter $V = 10^{13}$ and other parameters of the adaptive playback buffer policy to reasonable values. For the chosen parameters, we observe that the percentage of chunks which are skipped by the mobile user is $1.3 \%$ and the pre-buffering 
time is $162$ time slots. Furthermore, from Figure \ref{mobile-streaming}, showing the evolution of the playback buffer over time, 
we notice that there are no interruptions and the playback never 
enters the re-buffering mode.  
The helpers are numbered from $1$ to $25$, left to right and bottom to top, in Figure~\ref{topology}. 
In Figure~\ref{seamless}, we plot the helper index  providing chunk $k = 1, \ldots, 1000$ vs. the chunk index. 
We can observe that as the user moves slowly along the path, the DPP policy ``discovers'' adaptively 
the current neighboring helpers and downloads chunks from them in a seamless fashion. 
Overall, these results are indicative of the dynamic and adaptive nature of the DPP policy in response 
to arbitrary variations of large-scale pathloss coefficients due to mobility.  Extensive simulation results are presented in \cite{bethanabhotla2013joint} and are omitted due to space restrictions.

\section{Acknowledgement}
This material is supported in part by the Intel/Cisco VAWN program and by the
Network Science Collaborative Technology Alliance sponsored
by the U.S. Army Research Laboratory W911NF-09-2-0053.

\bibliographystyle{IEEEtran}
\bibliography{dilip-ref}

\end{document}